\documentclass[12pt]{revtex4}
\usepackage[english]{babel}
\usepackage{amsmath,amsthm}
\usepackage{amsfonts}
\usepackage{algorithmic}

\usepackage{graphicx}
\usepackage{textcomp}
\usepackage{algorithm}
\usepackage{setspace}

\usepackage{amsmath}  

\theoremstyle{definition}

\theoremstyle{remark}

\begin{document}

\title{A Fast Matrix-Completion-Based Approach for Recommendation Systems}
\author{MENG QIAO}
\affiliation{State Key Laboratory of Mathematical Engineering and Advanced Computing, Zhengzhou 45002, China}
\author{ZHENG SHAN}
\affiliation{State Key Laboratory of Mathematical Engineering and Advanced Computing, Zhengzhou 45002, China}
\author{FUDONG LIU}
\affiliation{State Key Laboratory of Mathematical Engineering and Advanced Computing, Zhengzhou 45002, China}
\author{XINGWEI LI}
\affiliation{State Key Laboratory of Mathematical Engineering and Advanced Computing, Zhengzhou 45002, China}
\author{WENJIE SUN}
\affiliation{State Key Laboratory of Mathematical Engineering and Advanced Computing, Zhengzhou 45002, China}

\author{LIEYI LUO}
\affiliation{Zhejiang NewBlue Network Media Co. Ltd., HangZhou 310000, China}\begin{abstract}
Matrix completion is widely used in machine learning, engineering control, image processing, and recommendation systems. Currently, a popular algorithm for matrix completion is Singular Value Threshold (SVT). In this algorithm, the singular value threshold should be set first. However, in a recommendation system, the dimension of the preference matrix keeps changing. Therefore, it is difficult to directly apply SVT. In addition, what the users of a recommendation system need is a sequence of personalized recommended results rather than the estimation of their scores. According to the above ideas, this paper proposes a novel approach named probability completion model~(PCM). By reducing the data dimension, the transitivity of the similar matrix, and singular value decomposition, this approach quickly obtains a completion matrix with the same probability distribution as the original matrix. The approach greatly reduces the computation time based on the accuracy of the sacrifice part, and can quickly obtain a low-rank similarity matrix with data trend approximation properties. The experimental results show that PCM can quickly generate a complementary matrix with similar data trends as the original matrix. The LCS score and efficiency of PCM are both higher than SVT.
\end{abstract}
\maketitle
\section{Introduction}
Matrix completion is a method for recovering lost information. It originates from machine learning and usually deals with highly sparse matrices. Missing or unknown data is estimated using the low-rank matrix of the known data~\cite{1}. This method is currently used in machine learning~\cite{2}, engineering control~\cite{3}, image processing~\cite{4}, and recommendation system~\cite{5}. In recommendation systems, it is difficult to obtain the overall preference matrix. Taking e-commerce as an example: since it is difficult for users to score each product, a large number of elements in the preference matrix are missing. Using the matrix completion method, we can predict the scores of unrated items based on the user's rating history~\cite{6}.

Singular value thresholding (SVT) is an algorithm for solving matrix completion problems\cite{5}. The main idea is to use the real data to optimize the kernel norm of the error matrix to obtain the optimal low-rank similarity matrix. The algorithm can obtain a more accurate low-rank similarity matrix, and has achieved satisfactory results in both image restoration and user scoring prediction~\cite{7}. However, the algorithm needs to set the singular value threshold during calculating. If the threshold is too large, a large amount of information will be lost. If it is too small, the running time of the algorithm will increase dramatically. Therefore, it is necessary to perform relevant experiments to acquire the optimal threshold setting. In recommendation system, the dimensions of the preference matrix and the singular values keep changing. Therefore, if a constant threshold is used, it may result in a poor recommendation.

There are other mature methods are used in recommendation systems. The most popular ones are based on matrix factorization. Matrix factorization methods are a type of collaborative filters, which can get the relationship between users and items. They have strong interpretability and good recommendation effect. The essence of these methods is to obtain the relationship between rows and columns by matrix decomposition. However, since the preference matrix in recommendation systems is generally large, it is difficult to achieve satisfactory results in a short time by directly applying decomposition. Therefore, most of the matrix factorization methods transform the decomposition process into a convex optimization problem by the gradient descent process~\cite{8}. However, the quality of the method is related to the amount of data. If the amount of data is large, the recommendation model cannot be updated in time. 

In this paper, we represent a light-weight recommendation system based on a fast matrix completion algorithm named probability completion model~(PCM). It quickly obtains a low-rank similarity matrix with the same probability distribution as the original matrix. As a result the recommendation system can quickly iterate according to the user's real-time clicks, capture and feedback the user preferences in time, and improve the user experience. In addition, we note that the users of recommendation systems care more about the sequence of final recommendations rather than their scores. Therefore, we use Longest Common Subsequence~(LCS) as the evaluation standard instead of F-norm or MAE. The LCS can focus measure the gap between the sequence~\cite{9}.
The main contributions of this work are as follows.

\begin{enumerate}

\item We propose an algorithm for recommendation systems, which is faster than the matrix factorization methods. 
\item We evaluate PCM using standard datasets. PCM achieves the same LCS score as the matrix factorization in a much shorter time.

\end{enumerate}

The remainder of the paper is organized as follows. Section II briefly overviews related studies and ideas on matrix completion. Section III presents our proposed approach in detail. Section IV describes experiments to validate our system and reports evaluation results. Section V summarizes our whole work and discusses several future directions.

\section{RELATED WORK}

This chapter mainly introduces the matrix completion algorithm and its related applications on the recommendation system.
\subsection{Matrix completion}

Matrix completion algorithm was initially applied to the problem of compressed sensing, proposed by Donoho, whose main goal is to obtain measurements of the original signal with fewer sampling resources~\cite{10}. At present, matrix completion algorithm technology has been applied to digital cameras, medical imaging, remote sensing imaging, and other fields~\cite{11}. Compressed sensing problems can be modeled as follows:
\begin{equation}
\min _{x \in R}\|X\|_{1} \quad \text { s.t. }  A x=y
\end{equation}
where $A$ represents pseudoinverse matrix, $x$ is a randomly sampled signal, and $y$ is the signal after completion.

An important premise of applying compressive sensing theory is the sparsity of signal vectors. However, the data used in many practical problems is a matrix rather than a one-dimensional vector. The sparseness of the matrix can be measured by the rank of the matrix. Therefore, the theory of compressed sensing in vector space naturally expands into the theory of matrix completion in matrix space. The matrix completion algorithm can be modeled as follows~\cite{12}:
\begin{equation}
\label{equ:rank}
\min _{x \in R} \operatorname{rank}(X) \quad \text { s.t.}\  P_{\Omega}(M)-P_{\Omega}(X) \leq \delta
\end{equation}
Where $P_{\Omega}$ is an orthogonal projection operator that projects sampling elements in range $\Omega$, i.e.:
\begin{equation}
P_{\Omega}(M)=\left\{\begin{array}{l}{M_{i j}, \text { if }(i, j) \in \Omega} \\ {0,\ \text  otherwise}\end{array}\right.
\end{equation}

Due to the non-convex and non-smoothness of the rank function, the standard matrix solution becomes an NP-hard problem \cite{13}. In 2002, Fazel proved that the kernel norm of the matrix can be used as the best convex approximation of the rank function~\cite{14}. Therefore, the formula~\ref{equ:rank} can be modified to obtain the following equation:
\begin{equation}
\min _{x \in r^{m \times n}}\|X\|_{*}\quad \text { s.t. } P_{\Omega}(M)=P_{\Omega}(X) 
\end{equation}

Through this transformation, the non-convex NP hard problem can be transformed into a convex constrained optimization problem. There are some classic methods, such as LASSO~\cite{15}, Bregman \cite{6}, and other linear iterative algorithms. However, these algorithms need singular value decomposition, which has high time complexity and is not suitable for large-scale matrix solving.

The methods based on kernel norm relaxation modeling involve the step of matrix singular value decomposition (SVD), and the time complexity of SVD is high. Therefore, Cai, Candes, and Shen \textit{et al.} proposed an SVT algorithm in 2008. The main idea is to use the low rank of the real data matrix. By optimizing the kernel of the error matrix, the optimal low-rank approximation of the existing data can be obtained. In the SVT algorithm, it is necessary to calculate all singular values whose matrices exceed a certain threshold before singular value decomposition. Because in each iteration step of the SVT algorithm the SVD of the data matrix is computed, the overall computational time of the SVT algorithm is significant. In 2017, Feng Wei \textit{et al.} proposed an SVT algorithm based on random singular value decomposition~\cite{7}, which improved the computational efficiency in the image recovery problem but not in recommendation systems.

\subsection{Matrix Factorization}

The Netflix Prize competition began in October 2006 and has played an important role in the development of Collaborative Filtering (CF). CF can provide personalized recommendations based on the user's behavior. The latent semantic analysis (LSA) is a type of CF~\cite{7}. The data types required by LSA include the user's purchase record, browsing history, search history, and even the movement of the mouse. LSA can explain the relationship between hidden features; therefore, the score can be predicted. Examples of popular LSA include the probabilistic latent semantic analysis (pLSA)~\cite{18}, the neural network~\cite{19}, the latent Dirichlet allocation~\cite{20}, and the SVD-based matrix factorization methods. Due to its accuracy and scalability, the matrix factorization methods are currently popular in recommendation systems.

\begin{center} 
\begin{figure}[t!]
\includegraphics[scale=0.5]{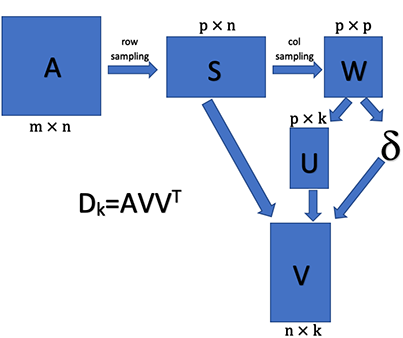}
\caption{Schematic diagram of PCM } 
\end{figure}
\end{center}

The use of matrix factorization methods requires the construction of the User-Item scoring matrix. However, in practice it is difficult to fully record the user's behavior, and it is difficult for the user to score all the items. The matrix information is incomplete and cannot be analyzed using traditional SVD. Therefore, the current popular method is to model the observed data.

The matrix factorization methods map the information of users and items into a joint latent semantic space of dimension $f$. Therefore, the user-item interaction can be modeled by the inner product in the space. The rules are as follows:
\begin{equation}
r_{u i}=\mu+b_{i}+b_{u}+q_{i}^{T} p_{u}
\end{equation}
where $b_{i}$ is the offset of item $i$, $b_{u}$ is user $u$, $q_{i}$ represents the vector mapped by item $i$ in $f$-dimensional space, $p_{u}$ represents the vector mapped by user $u$ in $f$-dimensional space, and $\mu$ represents the deviation from the model as a whole. At present, the mature matrix factorization methods are SVD, SVD++, timeSVD++, etc. Among them, SVD++ increases the user's degree of hobby modeling based on Equation 5:
\begin{equation}
p_{u}+|R(u)|^{-\frac{1}{2}} \sum_{j \in R(u)} y_{j}
\end{equation}
Where $R(u)$ contains all the items of the user $u$ score, and $y_{j}$ is the second item factor. These are all trainable parameters. 

TimeSVD++ builds user bias and item offsets into real-time functions that change over time. Taking the user offset $b_{u}$ as an example, the user offset can be expressed as:
\begin{equation}
b_{u}^{(1)}(t)=b_{u}+\alpha_{u} \cdot \operatorname{dev}_{u}(t)
\end{equation}
Where $b_{u}$ is the original user bias, $\alpha_{u}$ is the user's trainable parameter, and dev is the time offset function associated with the score. The expression is:
\begin{equation}
d e v_{u}(t)=\operatorname{sign}\left(t-t_{u}\right) \cdot\left|t-t_{u}\right|^{\beta}
\end{equation}
where $sign(x)$ represents the sign of $x$, $t_{u}$ represents the mean of the user's scoring time, and $\beta$ is the set value, which is generally set to 0.4 or 0.5.

Above methods can make a prediction of the items that the user has not scored. However, the accuracy of the recommendation result is related to the number of parameters. Therefore, matrix factorization methods are difficult to response and feedback in time.

\section{PROPOSED APPROACH}
It is difficult to use matrix completion algorithms or matrix factorization methods to generate recommendation results in time. Howver, in this section, we will present the PCM that can accelerate the calculation and capture user preference change in time. Section~\ref{sec:cert} and \ref{sec:rec} will introduce the PCM certification  and recommendation processes in detail.

\subsection{PCM process and proof}
\label{sec:cert}

Figure 1 shows a simplified workflow of PCM. The boxes represent matrices whose sizes are marked below. $A$ represents original matrix. $S$ and $W$ represent sampling matrix. $\sigma$ and $U$ represent singular values and singular vectors. $V$ is the result of multiplying $S$ by $U$, and $D$ is the completion matrix.

Using this method, the low-rank similarity matrix $D$ (acquired using matrix completion) of $A$ can be quickly obtained, so that the entire matrix can be quickly complemented. 

\begin{algorithm}
	\renewcommand{\algorithmicrequire}{\textbf{Input:}}
	\renewcommand{\algorithmicensure}{\textbf{Output:}}
	\caption{Probability Completion Model}
	\label{alg:1}
	\begin{algorithmic}[1]
		\REQUIRE latent dimension $K$, $G$, target predicate $p$
		\ENSURE $U^{p}$, $V^{p}$, $b^{p}$
		\STATE 	Let $A$ be a matrix of $m\times n$, and perform row sampling on $A$ to obtain a matrix $S$ whose size is $p \times n$, where $p$ is much smaller than $m$.
		\STATE 	Column sampling on the matrix $S$ to obtain a matrix $W$ of size $p \times p$
		\STATE 	Singular value decomposition on $W$ to obtain singular value matrix $U$ and corresponding singular value diagonal matrix $\sigma$
		\STATE  Taking the first $K$ singular values, the corresponding singular value matrix, and the matrix $S$ to obtain the matrix $V$, where $V \approx S^{T} U_{W}^{i} / \sigma^{i}$
		\STATE  Using the transposition of $V$ and $V$ to perform matrix multiplication with the original matrix $A$, the final complement matrix $D$ is obtained.
		\STATE \textbf{return} $D$
	\end{algorithmic}  
\end{algorithm}

As shown in Algorithm 1, The PCM only needs the matrix $A$ and the number of samples $p$ to quickly obtain the low-rank similarity matrix $D$. There is a similar data trend between $A$ and $D$. Therefore the element sequences of the two matrices are similar. Proof will be give below.

Let the original matrix be $A$, then $A$ can be expressed by the following equation:
\begin{equation}
\mathrm{A}=\sum_{t=1}^{r} \sigma_{t} u^{(t)} v^{(t)^{T}}
\end{equation}
Where $u$ is the left eigenvector of $A$ and $v$ is the right eigenvector of $A$. Let $D_{k}$ be the $k$-rank similarity matrix of $A$, then the formula of $D_{k}$ is as follows:
\begin{equation}
D_{k}=\sum_{t=1}^{k} A v^{(t)} v^{(t)^{T}}
\end{equation}

The derivation process of Equation 10 is as follows. Available from Equation 9
\begin{equation}
\mathrm{A} \sum_{t=1}^{r} v^{(\mathrm{t})}=\sum_{t=1}^{r} \sigma_{t} u^{(t)}
\end{equation}
Let there be a $k$-rank similarity matrix $D_{k}$ of $A$ ($k$ is less than or equal to the rank $r$ of $A$), then the expression is:
\begin{equation}
D_{k}=\sum_{t=1}^{k} \sigma_{t} u^{(t)} v^{(t)^{T}}
\end{equation}

For $D_{k}$, it can be written as $D_{k}=\sum_{t=1}^{r} \sigma_{t} u^{(t)} v^{(t)^{T}}-\sum_{t=k+1}^{r} \sigma_{t} u^{(t)} v^{(t)^{T}}$. Bring it into Equation 9 and 11, we can get $D_{k}=\sum_{t=1}^{r}Av^{(\mathrm{t})} v^{(t)^{T}} - \sum_{t=k+1}^{r} A v^{(\mathrm{t})} v^{(t)^{T}} =\sum_{t=1}^{k} A v^{(\mathrm{t})} v^{(t)^{T}}$.

It can be seen from Equation 10 that the $k$-rank matrix $D$ can be obtained from the right singular vector of the original matrix. However, since the size of the original matrix $A$ is too large, it is difficult to directly solve the SVD. Therefore, we can approximate the right singular vector $V$ of $A$ to approximate the $K$-rank matrix $D$.

In the original matrix, there is a trend in the whole data. We can sample the matrix by a certain probability sampling method, and reduce the data while retaining the trend of the data. This is equivalent to constructing a scaled-down matrix $S$. Based on the results of \cite{21}, we know that the matrix $S$ has the following relationship with the original matrix $A$:
\begin{equation}
S^{T} S \approx A^{T} A
\end{equation}
where $S$ is the row sample result of row sample $A$. $A$ and $S$ can be expressed as $A=U_{A} \xi V_{A}^{T}$, and $\mathrm{S}=U_{s} \xi V_{S}^{T}$, so that $A^{T} A=\sum_{i=1}^{r} \delta_{A_{i}}^{2} V_{A} V_{A}^{T}$ and $S^{T} S=\sum_{i=1}^{r} \delta_{S_{i}}^{2} V_{S} V_{S}^{T}$. The following is available from Equation 13:
\begin{equation}
V_{A} V_{A}^{T}=\sum_{i=1}^{r} \frac{\delta_{S_{i}}^{2}}{\delta_{A_{i}}^{2}} V_{S} V_{S}^{T}
\end{equation}

The right singular vector of $A$ available from Equation 13 can be approximated by the right singular vector of $S$. However the $S$ is also large. Therefore the data should be reduced again by sampling. Let $W$ be the column sample of $S$. According to Equation 13, we can get the following relationship between $W$ and $S$:
\begin{equation}
W W^{T} \approx S S^{T}
\end{equation}

It can be seen from Equation 14 that the left eigenvector of $W$ is similar to the left eigenvector of $S$. Therefore, the SVD of $S$ can be approximated by the left eigenvector of $W$ as
\begin{equation}
\mathrm{S} \approx U_{A} \Sigma_{S} V_{S}^{T}
\end{equation}
Then the right singular vector of $S$ can be approximated as
\begin{equation}
V_{S}^{i} \approx S^{T} U_{W}^{i} / \sigma^{i}
\end{equation}

Bringing Equation 17 and 14 into Equation 10, we can get the approximate solution $D_{k}$ of the $k$-rank matrix of $A$. However, the eigenvalues of matrix $A$ and matrix $S$ are used in Equation 14, which increases the computational complexity.

Since $S$ is the result of row sampling of $A$, linear transformation is performed on $A$ and $S$. According to the sampling rule, the $F$-norm $| | A| |_{F}$ of the $i$th row occupies the proportion of the overall matrix, and the sorting is performed from large to small. Let the transformed matrix be $\hat{A}$, $\hat{S}$, and the properties of the elementary linear transformation matrix invariant. Therefore, the eigenvalues of $A$ and $\hat{A}$, $S$ and $\hat{S}$ are the same. First, take the first row $\hat{A}_{1 \times n}$, $\hat{S}_{1 \times n}$ of $\hat{A}$ and $\hat{S}$. Since $\hat{S}$ is the sampling of $\hat{A}$, we get $||\hat{A}_{1 \times n}||_{F} \geq ||\hat{S}_{1 \times n}||_{F}$. The $F$-norm of a matrix has the following equation:
\begin{equation}
\|A\|_{F}=\sqrt{\sum_{j=1}^{m} \sum_{j=1}^{n}\left|a_{i j}\right|^{2}}=\sqrt{\operatorname{trace}\left(A^{*} A\right)}=\sqrt{\sum_{i=1}^{\min \{m, n\}} \sigma_{i}^{2}}
\end{equation}

It can be seen that the $F$-norm of the matrix is proportional to its eigenvalue. From Equation 18, we know that $\sigma_{\hat{A}_{1}} \geq \sigma_{\hat{S}_{1}}$. Then, go to the first two lines of $\hat{A}$ and $\hat{S}$, $||\hat{A}_{2 \times n}||_{F} \geq ||\hat{S}_{2 \times n}||_{F}$. $\sigma_{\hat{A}_{1}}\sigma_{\hat{A}_{2}} \geq \sigma_{\hat{S}_{1}}\sigma_{\hat{S}_{2}}$, since $\sigma_{\hat{A}_{1}} \geq \sigma_{\hat{S}_{1}}$, $\sigma_{\hat{A}_{2}} \geq \sigma_{\hat{S}_{2}}$. By analogy, $\sigma_{\hat{A}_{i}} \geq \sigma_{\hat{S}_{i}}$ can be obtained. Therefore, in Equation 14, $0 \leq \sum_{i=1}^{r} \frac{\delta_{s_{i}}^{2}}{\delta_{A_{i}}^{2}} \leq 1$, because $\delta_{A_{i}}^{2}, \delta_{S_{i}}^{2}$ is a descending sequence, and $\sigma_{\hat{A}_{i}} \geq \sigma_{\hat{S}_{i}}$, $\frac{\delta_{S_{i}}^{2}}{\delta_{A_{i}}^{2}}$ decreases. It can be seen that $V_{A} V_{A}^{T}$ is proportional to $V_{S} V_{S}^{T}$, and the overall data trend is the same. Setting $\asymp$ indicates that the data trend is the same. Therefore, according to the above and Equation 14, the following form is available:
\begin{equation}
V_{A} V_{A}^{T} \asymp V_{S} V_{S}^{T}
\end{equation}
Bring Equation 17, 19 into Equation 10, we can get the formula:
\begin{equation}
D_{k} \asymp \sum_{t=1}^{k} A v_{S}^{(t)} v_{S}^{(t)^{T}}
\end{equation}
$D_{k}$ has the same data trend as matrix $A$. For $\mathcal{V}_{S}^{(t)}$, there is $V_{S}^{i} \approx S^{T} U_{W}^{i} / \sigma^{i}$. Then, $v_{S}^{(t)} v_{S}^{(t)^{T}}=S^{T} U_{W}^{i} / \sigma^{i} \times U_{W}^{i^{T}} S / \sigma^{i}=S^{T} S / \sigma^{i^{2}}$.Since the number of rows after decimation is smaller than the rank of the original matrix $A$, the $S_{T}S$ is a real symmetric matrix and can be decomposed into the form of $\mathrm{U} \Sigma U^{T}$, where $U$ is an orthogonal matrix. Therefore, $v_{S}^{(t)} v_{S}^{(t)^{T}}=\mathrm{U} \Sigma U^{T} / \sigma^{i^{2}}$. The properties of the orthogonal transform are invariant. Therefore, $D_{k}$ has the same properties as the original matrix $A$, and their data trends are similar. Since PCM can get sequence results in time, it is more suitable for recommendation systems than classic matrix completion algorithms and matrix factorization methods.

\subsection{PCM recommendation process}
\label{sec:rec}

It is important to construct an appropriate preference matrix for PCM. In order to ensure the accuracy of the recommendation results, PCM requires that the dimensions of preference matrix do not change frequently. Therefore, when designing the matrix, it is necessary to adapt to the application. 

If PCM is used in e-commerce websites or movie websites, we can use the scoring matrix as the preference matrix since the number of goods or movies is fixed. Assume we use PCM in news websites or video websites. Because the updating rate of news and video is high, the matrix column update frequency will be high as well. Therefore, the scoring matrix does not have a fixed size. In this case, we need to abstract the scoring matrix.

Some items have higher frequency updates, which cannot directly constitute the user's scoring matrix. Instead, they need to abstract the item. For example, when building a news website, we can extract more fixed information such as keywords or title categories as an overview of the news, and then use these relatively fixed tags to construct the user's scoring matrix. When the news is scored, the relevant tags in the news are scored. The mapping between the tag and the news is modeled by the user reading interest model. This model is often referred to as a user profile~\cite{22}. It is possible to explore topics that may be of potential interest to the user. 

However, it is difficult to construct features or extract tags. Many small websites or companies do not have the resource to perform feature engineering. To avoid this problem, we can construct user profile through latent features. For example, in a video site, we can use the video co-views as preference matrix. This matrix provides a more abstract description of the user's preferences. It can also be considered as a matrix which scores between users. In this way, a group of users with a relatively high degree of association can be found.

\section{EXPERIMENTS}
\subsection{Methodology}
In order to measure the difference between the completed matrix and the real matrix, most matrix completion methods use Root Mean Squard Error (RMSE) or F-norm as the evaluation standard. However, in recommendation systems, the user care more about the recommendation sequence than items' scores. Therefore, instead of using RMSE or F-norm, we use longest common subsequence (LCS) as the evaluation standard. The calculation rules are as follows:
\begin{equation}
\text { rating }=\operatorname{LCS}\left(S_{1}, S_{2}\right) / \operatorname{len}\left(S_{1}\right)
\end{equation}
where $S_{1}$ represents the real matrix according to the scoring sequence, and $S_{2}$ is the scoring sequence obtained, e.g., by PCM. Since the total numbers of items are the same, the length of either $S_{1}$ or $S_{2}$ can be used. Rating is the final score with a value between 0 and 1. The difference between two sequences of the same length can be obtained by equation 21. The LCS can focus on the gap between the sequences. At present, LCS has been successfully applied to gene sequence comparison~\cite{23}.

Python3 is used as the programming language in the evaluation. The experimental platform is a Linux computer with 24G RAM and eight Intel Core\texttrademark I7-8550u CPUs running at 1.8 GHz.

The experiment is divided into two parts. The first part is the comparison between the PCM and the traditional matrix completion methods SVT. The second group is between the PCM and the matrix factorization SVD++. The experiment uses the movie-len small-scale dataset. The dataset contains a sparse matrix of $610$ users scoring $9,724$ movies. The matrix size is $610\times9724$, containing $100,936$ non-zero elements, which are scores in the 0.5\texttildelow5 range.

For SVT and PCM, we use 50\% elements of the dataset to make the sparse matrix. The test method is to compare the completed element sequence with another 50\% elemets' sequence. For SVD++, 50\% is extracted as the training set and 50\% is used as the test set.

The experiment performed a 10\% to 100\% ratio of the dataset to compare the recommended efficiency and effect in different data scales. The LCS score is used as the evaluation standard for the recommendation results.

For SVT, 100 iterations are performed. When the threshold is less than 0.0001, the iteration is terminated early. The K value is set to 30. For SVD++, 30 iterations were performed, and the K value was set to 50.

\subsection{Experimental Results}
The first is the comparision between the PCM and the SVT. It can be seen from Figure 2 that as the amount of data increases, The growth rate of PCM's time consumption is less than the SVT. We can see that the time of PCM with 100\% of dataset is similar to SVT with 10\% of dataset.

\begin{center} 
\begin{figure}[t!]
\includegraphics[scale=0.5]{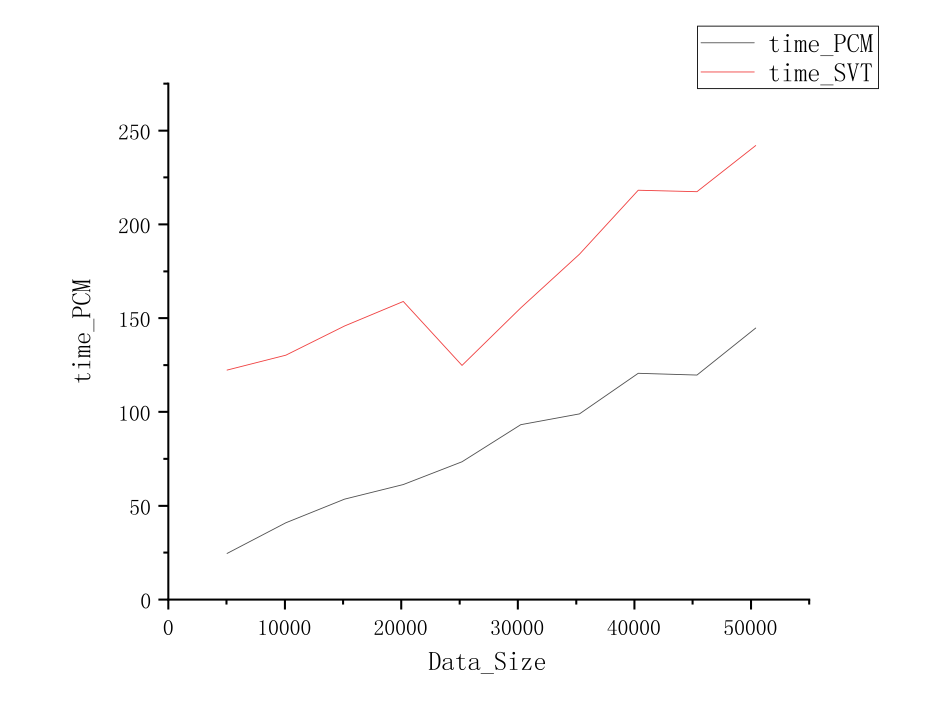}
\caption{PCM time comparison with SVT} 
\end{figure}
\end{center}

Next, the PCM is compared with the matrix factorization. Figure 3 shows the comparison of PCM and SVD++ over time in different datasets. It can be seen from this figure that as the data size continues increasing, the time consumption of PCM grows more slowly, while the time consumption of SVD++ follows an exponential rise.

\begin{center} 
\begin{figure}[t!]
\includegraphics[scale=0.5]{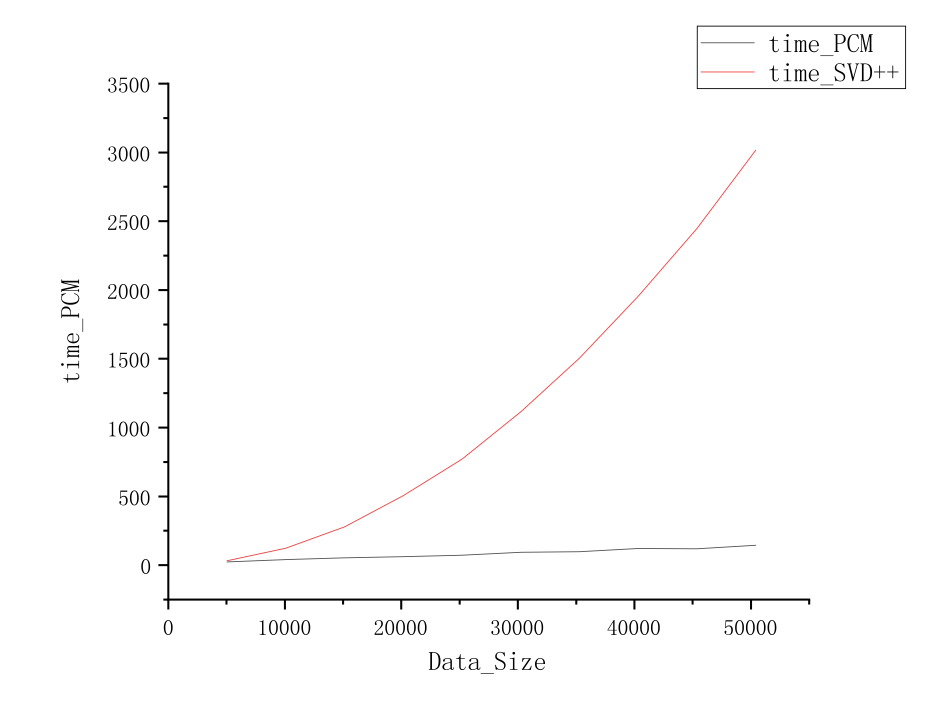}
\caption{PCM time comparison with SVD++} 
\end{figure}
\end{center}

Figure 4 is a comparison of LCS scores between PCM, SVT and SVD++ in different datasets. From this figure, we can see that under the different datasets, the overall LCS scores of the three methods are close. From Table 1, we can see that in recommended systems, SVD++ scores higher under different datasets, PCM scores second, and SVT scores are the lowest.

\begin{center} 
\begin{figure}[t!]
\includegraphics[scale=0.5]{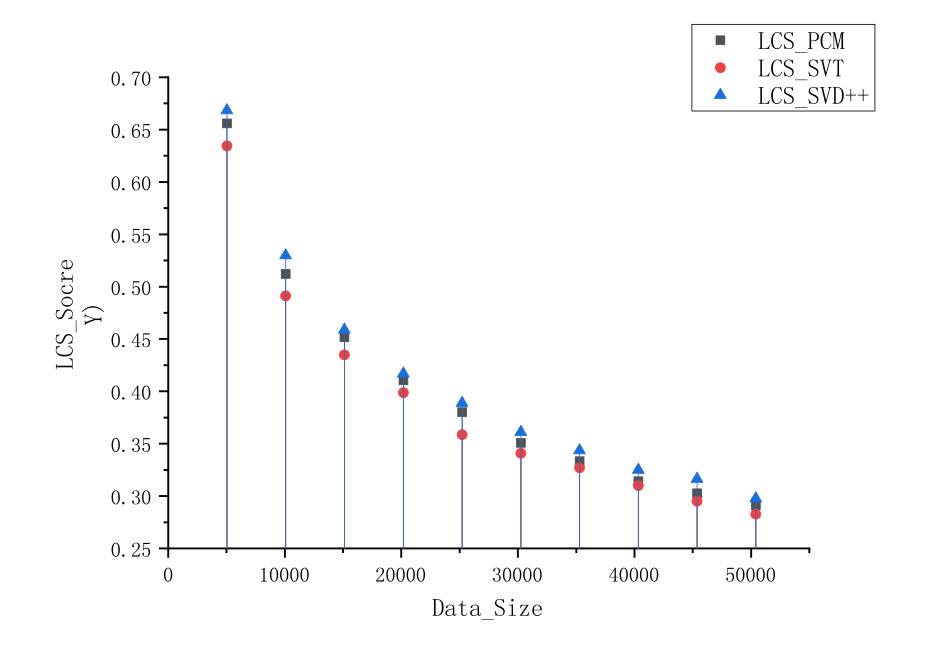}
\caption{PCM, SVD++ and SVT LCS score comparison} 
\end{figure}
\end{center}

Table 1 shows the detailed LCS scores of the three models of SVT, SVD++ and PCM under different datasets.  Figure 4 is a line graph of Table 1, from which it can be observed that PCM and SVD++ are almost coincident, and the SVT is slightly lower than the former two.

\begin{table}
\small
\centering
\renewcommand{\arraystretch}{1.3}
\caption{Units for Magnetic Properties}
\label{table}
\setlength{\tabcolsep}{3pt}
\begin{tabular}{c|c|c|c}
\hline
Data\_Size&LCS\_PCM&LCS\_SVT&LCS\_SVD++ \\ \hline
5041 &0.65604036&0.63447229&0.66831539 \\ \hline
10083&0.51219000&0.49128387&0.52974247 \\ \hline
15125&0.45158943&0.43490257&0.45880961 \\ \hline
20167&0.41065369&0.39880110&0.41676355 \\ \hline
25209&0.38029396&0.35878686&0.38882856 \\ \hline
30250&0.35090087&0.34099413&0.36096386 \\ \hline
35292&0.33332694&0.32698782&0.34344337 \\ \hline
40334&0.31410659&0.30994270&0.32486002 \\ \hline
45376&0.30276953&0.29507606&0.31622141 \\ \hline
50418&0.29124130&0.28279627&0.29769061 \\ \hline
\end{tabular}
\label{tab1}
\end{table} 
\subsection{Analysis}
It can be seen from the experimental results that SVD++ uses the gradient descent method to fit the parameters of the decomposed matrix through a large amount of data, which has certain advantages in accuracy and the recommended effect is better. However, due to the need for data for model training, as the amount of data increases, training time is also increasing. It is known from the principle of SVD that the time complexity of the PCM and SVT models depends on the size of the matrix, regardless of the amount of data. Therefore, when the amount of data increases, the PCM and the SVT do not have a large increase in time cost.

\section{CONCLUSION AND FEATURE WORK}
In this paper, we propose a probability completion model based on Monte Carlo SVD method, which can solve the matrix completion problem. PCM is suitable for recommendation systems. Compared with the classical matrix completion and the matrix factorization, PCM can perform the exponential acceleration while ensuring the accuracy of the recommendation effect. It can quickly capture changes in users' interest and give response in time.
However, the PCM needs to operate on matrix, therefore the memory consumption is a serious problem. We will use sparse matrix storage in next work. It can reduce the cost of memory in matrix operations.

\bibliography{ref}
\end{document}